\documentclass[pra,aps,showpacs,floatfix]{revtex4}
\usepackage{latexsym}
\usepackage{bm}
\usepackage{epsfig}
\usepackage{graphicx}
\usepackage{dcolumn}

\newcommand{\hpl}{${\rm H_2^+}$\ }

\begin{document}
\title{Visualizing quantum entanglement and the EPR
-paradox during the photodissociation of a diatomic molecule using
two ultrashort laser pulses}

\author{Szczepan Chelkowski, and
Andr\'{e} D. Bandrauk}
\affiliation{Laboratoire de Chimie
Th\'{e}orique, Facult\'{e} des Sciences, Universit\'{e} de
Sherbrooke, Sherbrooke Qc, J1K 2R1 Canada}

\begin{abstract}
 We investigate theoretically the dissociative
ionization of a \hpl molecule using two ultrashort laser
(pump-probe) pulses. The pump pulse prepares a dissociating
nuclear wave packet on an ungerade surface of \hpl. Next, an UV
(or XUV) probe pulse ionizes this dissociating state at large (R =
20 - 100 bohr) internuclear distance. We calculate the momenta
distributions of protons and photoelectrons which show a
(two-slit-like) interference structure. A general, simple
interference formula is obtained which depends on the electron and
protons momenta, as well as on the pump-probe delay on the pulses
durations and polarizations. This interference can be interpreted
as visualization of an electron state delocalized over the
two-centres. This state is an entangled state of a hydrogen atom
with a momentum $\vec p $ and a proton with an opposite momentum
$-\vec p$ dissociating on the ungerade surface of \hpl. This
pump-probe scheme can be used to reveal the nonlocality of the
electron which intuitively should be localized on just one of the
protons separated by the distance R much larger than the atomic
Bohr orbit.
\end{abstract}
\pacs{03.65.Ud, 82.53.Kp} \maketitle
\section{Introduction}

Recently, due to the extraordinary increase of research activities
in quantum information and quantum cryptography there is a growing
interest in various quantum intriguing phenomena originating back
to the famous Einstein-Podolski-Rosen (EPR) paradox
\cite{Einstein1935} formulated in 1935. This paradox is related to
the phenomenon of quantum entanglement \cite{Ekert2009} and to the
non-local character of quantum mechanics \cite{Dunningham07} or to
the problem of local realism versus the completness of quantum
mechanism. A most recent comprehensive review of various aspects
of the EPR paradox can be found in \cite{Reid2009}. So far nearly
all experimental evidence for entanglement is related to the
measurement of correlated photon pairs obtained in a process
called "parametric down-conversion \cite{Aspelmeyer2008}. In such
processes a photon from a laser beam gets absorbed by an atom
which subsequently emits two "polarization entangled" photons. The
entanglement phenomenon should in principle also appear in various
break-up processes involving slower (massive) fragments than
photons, e.g. in various disintegration processes such as
photoionization and photodissociation, as suggested in
\cite{Fedorov2006,Fedorov2004,Fry1995,Gneiting2008,Opatrny2001,Savage2007}.
So far, there exist very few experimental results demonstrating
the entanglement in such slower processes and involving massive
particles which are much better localized in space than massless
photons \cite{Reid2009}.

In this letter we investigate theoretically an experimental scheme
based on recent advances in the ultrashort laser technologies
which allow to shape laser pulses in femtosecond or even
sub-femtosecond (attosecond) time scales
\cite{Krausz2009,Vrakking2009}. Thus these technologies allow to
image the evolution in time of correlated electron-nuclear wave
function. This can be achieved by initiating the dissociation
process using an ultrashort pump pulse and allowing the
dissociating fragments to separate and be far apart. Next, this
system can be probed via a photoionization process using a probe
pulse having a well defined phase relative to the pump pulse, and
consequently in phase with the dissociating system. In general the
photoelectron spectra will exhibit a two-centre interference,
sometimes called a Fano interference since it was predicted by
Cohen and Fano in 1967 \cite{Cohen1966}. Their calculations showed
that when a molecule is photoionized via absorption of one photon
the photelectron spectra in diatomic molecules are modulated by an
interference factor
\begin{equation}
 \bar{\chi}=1\pm{\sin (|\vec{p}_e| R_{eq}) \over |\vec{p}_e| R_{eq}}\;\;,
\label{chifactor}
\end{equation}
 where $\vec{p}_e$ is the
electron momentum and $R_{eq}$ is the equilibrium internuclear
distance and the sign depends on the parity of the molecular
electronic wave function; (+) for a gerade and (-) for an ungerade
electronic state. We show that if the Fano interference is
observed in dissociating diatomic molecule at large internuclear
separation it may become an important tool for visualizing
peculiarities of quantum mechanics related to entanglement and to
the nonlocal character of the electron which (intuitively) should
localize on a single heavy (1836.15 times heavier than the
electron) centre, during the dissociation process, due to
localization via Coulomb attractive force.  Note that a quantum
state describing a localized electron on a specific proton
accompanying another distant proton would not lead to the
two-centre Fano interference in the photoelectron spectrum. This
interference is a result of the unique gerade or ungerade symmetry
of the electronic molecular wave function before the
photoionization takes place. Since the molecular Hamiltonian has
also this symmetry we conclude that the state before the turn-on
of the probe pulse should preserve the symmetry of the
dissociating state prepared by a pump pulse. In other words,
quantum dissociating state is not a simple product of the hydrogen
and proton states but is a coherent superposition of the possible
"simultaneous presence" of the electron on both well separated
protons. Thus in this specific dissociation process we face
clearly the conflict with local realism: in any deterministic
theory the simultaneous presence of the electron on two heavy well
separated protons should not occur, and, consequently the
observation of Fano interference pattern is in conflict with local
realism in a similar way as the experimental violation Bell
inequalities is \cite{Reid2009}.\\
\begin{figure}[!ht]
\includegraphics[width=0.8\textwidth]{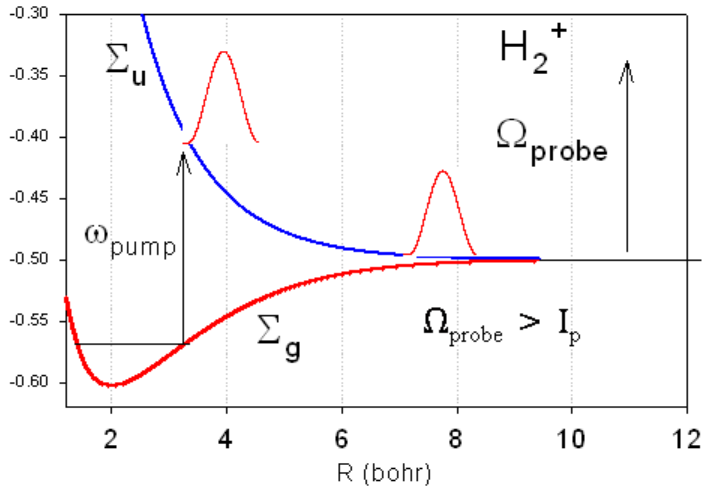}
\caption{Illustration of the proposed pump-probe experimental
scheme. Two lowest electronic surfaces $\Sigma_g$ and $\Sigma_u$
of \hpl are shown. The pump pulse prepares the nuclear wave packet
sliding on the upper ungerade surface $\Sigma_u$.  After the
turn-off of the pump pulse this wave packet evolves as free system
until it is photoionize at large internuclear distance R. }
\end{figure}
More specifically, we consider a pump-probe excitation scheme
\cite{Krausz2009,Vrakking2009} in which a pump laser pulse
prepares a dissociating nuclear wave packet on an ungerade
(repulsive) surface of a \hpl molecule. Next (20-200 fs later,
after the pump pulse is turned-off), a UV (or XUV) probe pulse
ionizes this dissociating (H-atom + proton) state at large (R = 20
- 150 bohr) internuclear distance, as illustrated in Fig.1 and
also described in \cite{Vrakking2009}. We show that coincidence
measurement of the electron and proton spectra reveal a very
special, counter-intuitive nature of the quantum dissociation
process. It can be assumed that after the turn-off of the pump
pulse the motion on the ungerade surface of \hpl is adiabatic.
Consequently, because the antisymmetric (or symmetric if
disociation occurs on a gerade electronic state) character of the
electronic wave function the quantum state of the dissociating
system is very distinct from a state of a free proton and a free
hydrogen atom \cite{Bykov2003} which for instance occurs in
proton-hydrogen scattering. Thus quantum mechanics predicts that
even at large internuclear distance the electron can be well
localized in two places, i.e. it can form a hydrogen atom on two
well separated protons which seems counter-intuitive. Note that in
our scheme electron localization occurs due to the Coulomb
attraction from two protons. Suppose that we have measured the
hydrogen atom at the right hand side along the laser polarization
vector, as shown in Fig.2 . Already at relatively
small internuclear distance R=10 bohr shown in Fig.2 the electron
is very well localize on each centre. Thus we infer (using the charge
conservation principle) that the opposite detector (at left-hand
side in Fig.2) at the internuclear separation larger that R=60
bohr will measure with a certitude a proton. This situation
resembles a variant of the EPR paradox based on the disintegrating
system of the two, spin one-half particles, originating from
their initial singlet state \cite{Sakurai1985} in which by
measuring a spin up by one detector we infer what was the spin
projection measured by the opposite detector.\\
\begin{figure}[!hb]
\includegraphics[width=0.8\textwidth]{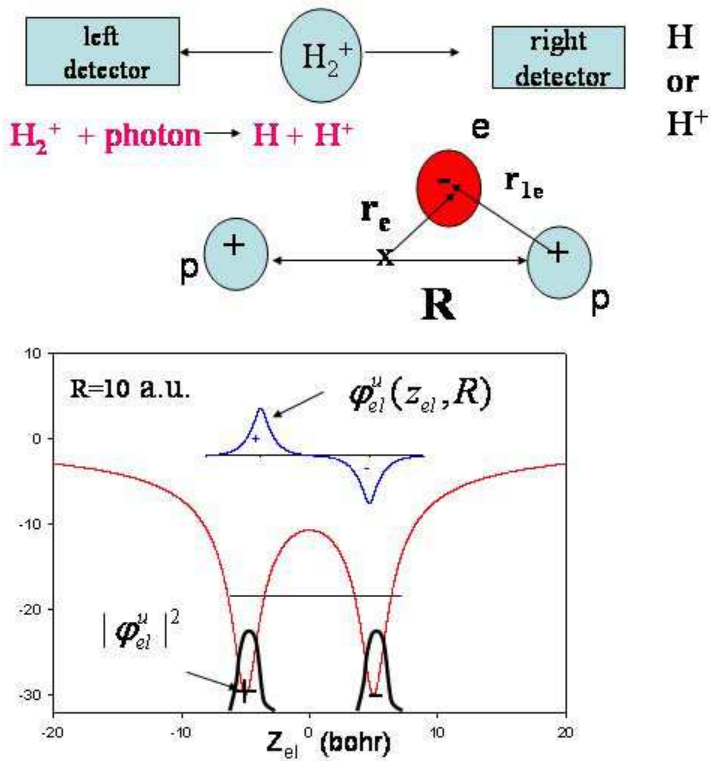}
\caption{The upper part shows schematically the first stage of the
proposed experimental scheme in the case when only the laser pump
pulse is used. This pulse dissociates the \hpl molecule. Two
opposing detectors are measuring the H-atom or the H$^+$ ion .
These measurements resemble the case of the EPR paradox in which
the spin projections are measured in the disintegration of the two
spin one-half particles from the initial single spin state. We
also illustrate the definition of the Jacobi coordinates for the
p-p-e system used in our calculations. The lower part show the
two-centre Coulomb potential, the ungerade electronic wave
function $\varphi_{el}^u$ and the corresponding probability
density as function of the electron coordinate $z_{el}$ at fixed
intenruclear distance R=10 bohr.}
\end{figure}
Suppose now that instead of measuring directly the H-atom in the above experiment we use
a probe UV pulse which photoionizes the molecule and
the two-centre Fano interference
pattern is observed in the photoionization signal.
Simple perturbative calculations using a plane wave
approximation for the final electrons state
\cite{Cohen1966,Yudin2006} predict that at fixed internuclear
vector $\vec R $ the molecular photoelectron signal is modulated
via interference factors $\sin^2(\vec p_e \cdot \vec {R}/2)$ or $\cos^2(\vec
p_e \cdot \vec {R}/2)$.
Just the fact that we observe
such an interference at large $|\vec{R}|$ would be
a witness of the two following interesting
quantum phenomena.
First, in the proposed experiment the probe pulse will monitor an
entangled state of two particles: the hydrogen atom and the proton,
represented by the state $|H, \vec p >$ times and the $H^+$ (i,e, proton)
flying in an opposite direction represented by a ket $|H^+, -\vec
p >$. The dissociating state is not a simple product of these basis states
but because of the electron symmetry of the state dissociating on
a single electron surface we must add to the product: $|H,
\vec p > |H^+, -\vec p >$  the product $|H, -\vec p
> |H^+, \vec p >$ and integrate over the relative momentum $\vec p$
weighed with the momentum distribution of the dissociating wave
packet. If a two-centre interference pattern is observed in the
photoelectron signal this means that we observe an entangled
state, since a simple product of these states cannot yield a two
centre interference pattern. Second, any local hidden variable
theory would predict the electron position with respect to one of
the protons is well defined within the Bohr radius. Thus when the
interference pattern is observed in the photoelectron spectrum
this means that the electron is well localized simultaneously on
both well separated protons before we photoionized this quantum
system. This is a very unusual situation since the tunneling time
from one centre to another is extremely  large even at the
relatively small internuclear separation R=60 bohr: t$_{tunnel}$=
2.4 years \cite{Bykov2003}. Moreover, if we interpret the integral
of e$|\psi |^2$ over half-space as a charge present around a
specific proton one concludes that a fractional charge e/2 is well
localized around one centre \cite{Bykov2006}. Thus one may argue
that the observation of the Fano two-centre interference is a
witness for the simultaneous presence of a charge e/2 on each
centre. Clearly, this simple 3-body system with the electron and
two well separated protons represents an interesting quantum
mystery related to the formation of a hydrogen atom during the
dissociation process and is certainly worth further experimental
and theoretical investigation.

In our theoretical description of the above mentioned pump-probe
scheme (dissociation followed by photoionization), we do not
calculate the dynamics of the first step in which an ultrashort UV
pump pulse photodissociates (via an absorption of one photon) the
\hpl molecule and it thus prepares a nuclear wave packet moving on
the molecular ungerade surface. We assume that this wave packet
has the Gaussian shape right after the pump pulse is turned off,
centres at $R$=$R_0$=12 bohr, at t=0, and it
 next evolves as a free system (field-free) on
the $\Sigma_u$ surface up to R=60-120 bohr when an ultrashort UV pump
laser pulse is turned-on and
it photoionizes this dissociating packet also via an one photon process.
The photoionization
probability distributions of the momenta of the electron and the protons
are calculated using first order perturbation theory in 3-D
(3-D for both electron and nuclear degreees of freedom).
A plane wave approximation is used for the final state of
protons and a "modified-plane-wave" approximation for the electron (see next
Section for the details).
Note that to the best of our knowledge, all theoretical work
related to Fano interference has been so far done using frozen nuclei at
fixed equilibrium internulear distance $|\vec{R}_{eq}|$.
We believe that our study is the first dynamical investigation of
Fano interference in the photoelectron spectrum originating from
dissociating molecules at large internuclear separations. It includes
full 3-D electron-nuclear dynamics. Note, that so far approaches based on solving
time-dependent Schr{\"o}dinger equation for \hpl are based on the models with
reduced dimensionality \cite{Bandrauk2009,Bandrauk1999}.

\section{Perturbative calculations of dissociative-ionization}

We use Jacobi coordinates for the electron and two protons
(separated by a vector $\vec R$) in which the electron position
vector $\vec r_e$ originates in the centre of mass of two protons,
see Fig.2. In these coordinates the Hamiltonian of \hpl has the
following form \cite{Hiskes1961}: (in atomic units,
$\hbar=m_e=e$=1)
\begin{equation}
\hat{H}_{0}= -{1 \over 2m'_e} \Delta_{\vec r_e}-{1 \over 2\mu}\Delta_{\vec
R}+V_C(\vec r_e,\vec R) \; , \label{hamilt1}
\end{equation}
where $$m'_e =\; {2m_p m_e \over 2m_p+ m_e}\;, \mu=m_p/2$$ are the
electron - two proton reduced mass and the proton-proton reduced
mass, respectively,
\begin{equation}
V_C(\vec r_e,\vec R)\;=\;{-1 \over |\vec r_e-\vec R/2|}\;+ \;{-1
\over |\vec r_e+\vec R/2|}\;+\;{1 \over |\vec R|} \label{coulomb}
\end{equation}
is the total Coulomb interaction between protons and the electron and
$m_e$ and $m_p$  are the electron and proton masses.
The total Hamiltonian is $\hat{H}=\hat{H}_0+\hat{V}_{int}$ where
\begin{equation}
\hat{V}_{int}=-i{\kappa \over m'_ec} \vec A(t) \nabla_{\vec{r}_e},
\label{vint}
\end{equation}
where $$\kappa\;=\;1+{m_e \over {2m_p+m_e}}\;\;.$$
$V_{int}$  describes the
interaction of \hpl with a UV laser probe field via its vector
potential. Since we use a weak
intensity probe pulse and the pulse frequency is larger
than ionization potential of \hpl we may use the perturbation theory
for calculation of the photoionization probability of the dissociating
wave packet (prepared by the pump pulse) using
the transition amplitude:
\begin{equation}
A_{fi}
=-i\int_{-\infty}^{\infty}dt\langle\psi_{f}|e^{i\hat{H}_{0}t}V_{int}
e^{-i\hat{H}_{0}t}|\psi_{in}(t)\rangle  \label{amplitude}
\end{equation}
We first recall the
result for fixed nuclei at $\vec{R}$. Using for the electron
initial state
\begin{equation}
\varphi_{el}^{g/u}(\vec{r}_e,\vec{R})
=\frac{1}{\sqrt{2}}\left[\psi_{H}(\vec{r}_{e}+\vec{R}/2)
\pm \psi_{H}(\vec{r}_e - \vec{R}/2)\right]
\label{psiel}
\end{equation}
where $\psi_H$ is the hydrogen 1s wave function (sign ($\pm$) is
used for the electronic gerade on ungerade state), and using the
plane wave for the final electron state:
\begin{equation}
\psi_{f}=(2\pi)^{-3/2}\exp(i\vec{p}_{e}\cdot\vec{r}_{e})
\label{planewave}
\end{equation}
we get from eq. (\ref{amplitude}) (just by choosing the
integration coordinates local to each centre) that
\begin{equation}
|A_{fi}^{fix}|^2\;\sim \; \left[1\pm \cos(\vec p_e\cdot \vec R)\right] |A_H(\vec p_e)|^2
\label{fixed}
\end{equation}
where $A_H(\vec p_e)$ is the atomic photoionziation amplitude given in
\cite{Yudin2006}. Thus if ionization occurs from the ungerade surface
the
molecular ionization probability is modulated via a $\sin^2(\vec p_e
\cdot \vec R/2)$ factor. If the molecule is not initially aligned,
and if protons momenta are not measured, we need to integrate over
the direction of $\vec{R}$ which
leads to the $\bar{\chi}$ factor given in eq.(\ref{chifactor}). To
include the nuclear motion in eq.(\ref{amplitude}) we use the plane
wave approximation for the final state for both the electron and
for the relative motion of protons:
\begin{equation}
\psi_{f}=(2\pi)^{-3}\exp(i\vec{p}_{N}\cdot\vec{R})
\exp(i\vec{p}_{e}\cdot\vec{r}_{e}).
\end{equation}
We will isolate from this expression the electronic atomic
photoionization amplitude $A_H(\vec{p}_e)$ in which we will use
the exact expression for an atomic amplitude, i.e. in which exact
Coulomb wave for the individual centre is included (but the
influence of the neighboring centre is neglected which is a
reasonable approximation for the case of very large R we are
interested in). This explains the term "modified-plane-wave"
approximation used in the introduction. Regarding the plane wave
approximation for the nuclei we think that it is justified for
photoionization occurring at very large internuclear distances. We
are interested in R's as large as $R>60$ bohr at which Coulomb
repulsion should be negligible with respect to the kinetic energy
of the the dissociating \hpl , i.e. when
\begin{equation}
\frac{p_{0}^{2}}{2\mu}\gg\frac{e^{2}}{R_{0}+p_{0}t_{c}/\mu}\;\;
\end{equation}
where $R_0$ is the position of the the centre of the dissociating wave packet
with the momentum $p_0$ at t=0 when the pump pulse is turned off and $t_c$ is
the time at which the amplitude of the probe pulse is at maximum. At this time
the wave packet reaches the distance $R_0+p_{0}t_{c} / \mu\;$.
These plane wave approximations allow us to get a simple analytic expression
for the amplitude $A_{fi}(\vec{p}_e,\vec{p}_N,\tau,\Delta R,t_c)$.
Using the Coulomb waves in the final state leads to much more complicated
formulas involving some numerical integration.

As the initial state we take the Born-Oppenheimer
solution (we consider an improvement to this approximation
in the appendix)
as a product of the ungerade electronic function (\ref{psiel}) and
the superposition of nuclear plane waves $\exp(i \vec{p}\cdot \vec R)$ :
\begin{equation}
\psi_{in}( \vec{r}_e,\vec{R},R_0)=\varphi_{el}^{u}(\vec{r}_e,\vec{R}) \int d^{3}p
\; \varphi_{N}(\vec p,R_0)
\exp (i\vec{p}\cdot \vec{R})
\label{initialbo}
\end{equation}
where $\varphi_{N}(\vec p, R_0)$ is the initial distribution
of the momenta in the dissociating nuclear wave packet. It should
be adjusted to the shape of the wave packet prepared by the pump pulse
which we suppose is short and in the amplitude (\ref{amplitude})
we assume the free evolution of the \hpl wave packet between
the turn-off the pump pulse at t=0 and the turn-on of the probe pulse.
We derive next the analytic formula for photoionization valid
for any shape of $\varphi_{N}(\vec p,R_0)$. Its specific shape
will be chosen for illustrating graphically our results.
In order to perform analytically the time integral
in (\ref{amplitude}) we need a specific shape of the vector
potential $\vec A(t)$ of the laser field. We assume it has a Gaussian form:
\begin{equation}
\vec{A}(t)=\frac{\vec{e}_{probe}}{2}A_{0}\exp\left[-\frac{(t-t_{c})^{2}}
{2\tau^{2}}\right]\exp(-i\Omega_{probe}t)+C.C.\;\;
\label{pulse}
\end{equation}
where $A_0$, $\vec{e}_{probe}$, $\Omega_{probe}$, $\tau$ are the
UV probe pulse amplitude, polarization, its central frequency and
duration. Pulse duration $\tau$ is related to the commonly used FWHM
duration via relation $\tau_{FWHM}=2 \sqrt{\ln(2)} \tau$. We recall
that thus defined FWHM means full width at half maximum of the
laser intensity time profile, not the FWHM of the envelope of the
laser field. $t_c$ is the time at which the probe pulse has maximum
and at the same time  $t_c$ is also a measure the time delay between
the probe and the pump pulse since we have chosen t=0 as time when
the pump pulse is turned off and the centre of the nuclear packet is
at $R=R_0$.

After integrating
the time t and the electronic coordinate $\vec{r}_{e}$, in the
formula (\ref{amplitude}) we get :
\begin{equation}
A_{fi}=N_{1}A_{H}\left(\vec{p}_{e}\right) \int d^{3}p
\;a(\vec p)\; \int d^{3}R
\exp \left(i\vec{p}\cdot\vec{R} -i\vec{p}_{N} \cdot
\vec{R} \right)
\left(e^{i\vec{p}_{e}\cdot\vec{R}/2}
-e^{-i\vec{p}_{e}\cdot\vec{R}/2}\right)
\end{equation}
where
\begin{equation}
a(\vec p)=\exp [if(\vec p) t_{c}-\tau^{2}f(\vec{p})/2]
\varphi_{N}(\vec{p},R_0) \;\;,
\label{adef}
\end{equation}
\begin{equation}
f(\vec p)={\vec {p}\;^2_e \over 2m'_e}+{\vec {p}\;^2_N \over
2\mu}+I_p-\Omega_{probe} -{\vec {p}\;^2 \over 2\mu}\;\;,
N_1={\kappa A_0 \tau \over 2\pi 2^{3/2} m'_e  c}\;,\;
\end{equation}
\begin{equation}
A_{H}(\vec{p}_{e})=A_{1s}(\vec{p}_{e})
=2^{3/2}\frac{(\vec{e}\cdot\vec{p}_{e})(1-i\nu)}{\pi(1+\vec{p}_{e}^{2})^{2}}
\exp\left[-2\nu\arctan(p_{e})\right]N_{\nu}^{*}\;,
\label{atomic}
\end{equation}
$$\nu={1 \over p_e}\;\;,\;\;N_{\nu}=\exp(\pi\nu/2)\Gamma (1+i\nu)\;\;\;.$$
Integration over the $\vec R$ coordinate yields two Dirac delta functions
for momentum conservation. Thus
we get for the dissociative-ionization amplitude
\begin{equation}
A_{fi}=N_{1}(2 \pi)^{3}A_{H}(\vec{p}_{e})\int
d^{3}p\;a(\vec{p})\;
[(\delta(\vec{p}-\vec{p}_{N}+\vec{p}_{e}/2)
-\delta(\vec{p}-\vec{p}_{N}-\vec{p}_{e}/2)]
\end{equation}
Integration over the nuclear momenta $\vec p$ yields
the final probability amplitude for dissociative-ionization
\begin{equation}
A_{fi}(\vec{p}_e,\vec{p}_N,\tau,\Delta R,t_c)
=N_2 A_{H}(\vec{p}_{e})\left[ a(\vec{p}_{-})-a(\vec{p}_{+})\right]
\label{amplgeneral}
\end{equation}
where $a(\vec p)$ is defined in (\ref{adef}), $$N_2=(2\pi)^3
N_1={\kappa A_0 (2\pi)^2 \tau \over 2^{3/2} m'_e c} \;\;, $$ and
\begin{equation}
\vec{p}_{\pm}= \vec{p}_{N} \pm \vec{p}_{e}/2\;\;. \label{plmi}
\end{equation}
The shifted momentum in the last equation is related to the recoil
received by each proton
from the electron: the final relative momentum is either $\vec {p}_{+}$ or
$\vec {p}_{-}$.

Eq.(\ref{amplgeneral}) provides a general expression for the momenta distributions
of protons and of the electron valid for any initial distribution of momenta
$\varphi_{N}(\vec{p},R_0) $ of a dissociating wave packet. However,
in order to investigate in detail Fano two-centre interference effect
we need to specify the initial momentum
distribution in the wave packet prepared by the pump pulse. We
suppose that the \hpl molecule was initially in the vibrational
v=0, J=0 of the gerade bound state, where J is the initial angular
momentum quantum number of the molecule. Thus the nuclear wave packet,
after absorption of one photon, will be in
the J=1 rotational state on the ungerade $\Sigma_u$ surface of
\hpl . We assume that it has the form:
\begin{equation}
\varphi_{N}\left(\vec{p},R_0\right)=
C_{N}\frac{\cos\theta_{p}}{p}\exp\left(-\frac{\Delta
R^{2}}{2}(p-p_{0})^{2}\right)\exp\left[i(p_{0}-p)R_{0}\right]
\label{initialp}
\end{equation}
where $\cos\theta_{p}=\vec{p}\cdot\vec{e}_{pump}/|\vec{p}|$ and
$\theta_{p}$ is angle between the nuclear relative momentum
$\vec{p}$ and the pump pulse polarization. The nuclear wave packet
is at $R=R_{0}$ at $t=0$, the probe has maximum at $t=t_{c}$. This
is a free Gaussian wave packet (in the radial variable
$p=|\vec{p}|$ ) sliding on the electronic surface $\Sigma_u$ (its
electronic wave function $\varphi_{el}^{u}$ is given in
(\ref{psiel}) with the angular momentum J=1. The central radial
momentum of the wave packet is $p_{0}$ and its spatial width is
$\Delta R$. In order to study the two-centre interference effect
it is convenient to rewrite the probability of ionization in the
following form (in this form the interference appears through the
cross term C):
\begin{equation}
|A_{fi}(\vec{p}_e,\vec{p}_N,\tau,\Delta R,t_c)|^2=
|A_H(\vec{p}_e)|^2 \left[|a(\vec {p}_{-}|^2 + |a(\vec {p}_{-}|^2
+C({p}_{+},{p}_{-},t_c) \right]
\label{probab}
\end{equation}
where
\begin{equation}
|a(\vec p)|=C_N \exp \left(-f^2(\vec
p)\frac{\tau^2}{2}-\frac{\Delta R^{2}}{2}(p-p_{0})\right) |\vec p
\cdot \vec e|\;/|\vec p |^2, \;\;,
\end{equation}
\begin{equation}
C({p}_{+},{p}_{-},t_c)=2|a(\vec {p}_{+})|\;|a(\vec {p}_{-})|
\cos (\Phi (t_c))\;\;,
\label{interfc}
\end{equation}
\begin{equation}
\Phi (t_c,\vec{p}_e\;,\vec{p}_N)= (|\vec {p}_{+}|-|\vec{p}_{-}|)R_0
+ \left[f(\vec {p}_{-})|- f(\vec {p}_{+})\right]t_c=
(|\vec {p}_{+}|-|\vec{p}_{-}|)R_0+{|\vec {p}_{+}|^2-
|\vec {p}_{-}|^2 \over 2 \mu}\;t_c \;.
\label{Phi}
\end{equation}
After the use of eq.(\ref{plmi}) the phase $\Phi$ becomes
\begin{equation}
\Phi (t_c,\vec{p}_e, \vec{p}_N)= (|\vec {p}_{+}|-|\vec{p}_{-}|)\;R_0
+\vec{p}_e \cdot {\vec{p}_N\over \mu}\;t_c\;\;.
\label{Phin}
\end{equation}
This is an important result since after comparing
eq.(\ref{Phin}) with the phase corresponding
to static result eq.(\ref{fixed})
we conclude that by measuring the relative nuclear momentum
$p_N$ and the delay time $t_c$ we are fixing the increment of
the internuclear separation $\vec{R}$ during the time interval $t_c$, i.e.
this increment is simply:
$\vec{v}_N t_c$ where $\vec{v}_N=\vec{p}_N \; / \;\mu$ is the
relative velocity of protons.
We can simplify more equation (\ref{Phin}) in the case of
the nuclear momentum larger than the electron momentum, i.e. if
the inequality $ p_e\; << \; p_N $ holds we get:
\begin{equation}
|\vec {p}_{\pm}|=\sqrt{p_N^2 \pm \vec p_e \cdot \vec p_N+p_e^2/4}
\simeq p_N \pm \vec p_e \cdot {\vec p_N \over 2p_N}\;=\;p_N \pm
p_e \cos ( \theta_{pe})\; /2;\;. \label{pmapprox}
\end{equation}
Thus the phase (\ref{Phin})
on which the interference relies simplifies to:
\begin{equation}
\Phi (t_c,\vec{p}_N))\simeq \vec{p}_e\cdot
\vec{R}_N(t_c,\vec{p}_N) \;\;, \label{Phiappr}
\end{equation}
where
\begin{equation}
\vec{R}_N(t_c,\vec{p}_N)\simeq R_0{\vec{p}_N \over
|\vec{p}_N|}+{\vec{p}_N t_c \over \mu}\;\;. \label{rtcappr}
\end{equation}
We will also use later the absolute value of the vector $\vec{R}_N$:
\begin{equation}
R_N(t_c,p_N)=|\vec{R}_N(t_c,\vec{p}_N)|\simeq R_0 +{p_N t_c \over
\mu}\;\;. \label{rtcapprab}
\end{equation}

Clearly, we see from the last equations that, as in the case
of static result (\ref{fixed})
the interference term $C({p}_{+},{p}_{-},t_c)$ is
modulated via the term
$ \cos \left[ R_N(t_c,p_N)p_e \cos (\theta_{ep})\right]\;,$
where $\theta_{ep}$ is the angle between the electron momentum
and relative nuclear momentum $\vec{p}_N$. This relation can be
used for imaging the nuclear motion as suggested in \cite{Vrakking2009}:
if we measure the ionization signal for a series of time delays $t_c$ and
follow the change of a specific minimum in the spectrum, we can thus
deduce the molecular trajectory from the relation $2n\pi /p_e cos(\theta_{ep})$, where
n is an integer corresponding to a specific minimum.
\begin{figure}[!hb]
\includegraphics[width=0.8\textwidth]{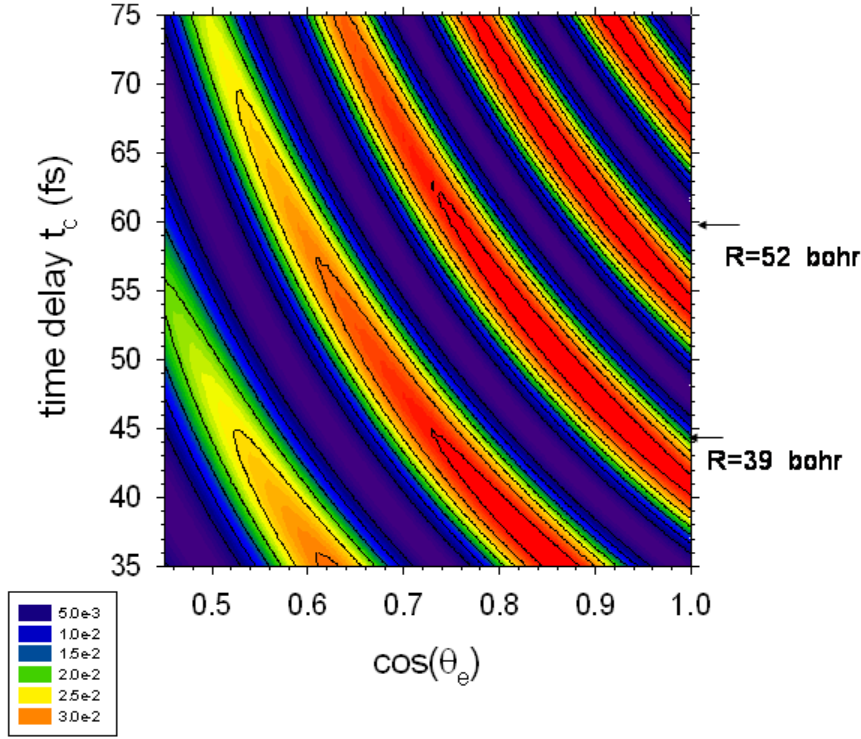}
\caption{Ionization probability calculated using eqs.(21)-(24) for
the case when the polarizations of the pump and probe pulses are
parallel and the nuclear relative momentum $\vec{p}_N$ is also
parallel to both polarizations. We used: $\lambda_{probe}$=60 nm,
$p_N=p_0=$14.8 a.u.,$p_e$=0.72 a.u., $\Delta R$= 3.0 bohr,
$R_0$=12.0 bohr and the pump pulse duration $\tau_{FHWM}$=2.4 fs.}
\end{figure}
If furthermore, the width of the momentum distribution ${1 \over
\Delta R}$ is sufficiently large compared to the electron momentum
$p_e$, i.e. $p_e < {1 \over \Delta R}$, we may expect that the
following approximations are valid for $p_N$ values close to the
central value $p_0$ of the momentum distributions defined via
(\ref{initialp}):
\begin{equation}
|a(\vec {p}_{-}| \simeq |a(\vec{p}_{+}|)\;\simeq |a(\vec{p}_{N}|)\;\;,
\label{approxa}
\end{equation}
we get
\begin{equation}
|A_{fi}|^{2}\sim |A_{H}(\vec{p}_{e})|^{2}
\sin^{2}\left(\vec{p}_{e}\cdot\frac{\vec{p}_{N}}{|\vec{p}_{N}|}R_N(t_{c}, |\vec{p}_N|)
/2\right)\left[{(\vec p_N \cdot \vec{e}_{pump}\over p_N^2}\right]^2
\end{equation}
or
\begin{equation}
|A_{fi}|^{2}\sim \; \cos^2(\theta_{e})\sin^{2}\left[ p_{e}
\cos\theta_{pe}
R_N(t_{c},|\vec{p}_N|)/2\right]\cos^{2}(\theta_{p}) \label{fanopn}
\end{equation}
where $\theta_e$ is the angle between the electron momentum and
the probe pulse polarization vector $\vec e_{probe}$, and $\theta_{p}$ is
the angle between the $\vec p_N$ vector and the pump polarization $\vec e_{pump}$ .
\begin{figure}[!hb]
\includegraphics[width=0.8\textwidth]{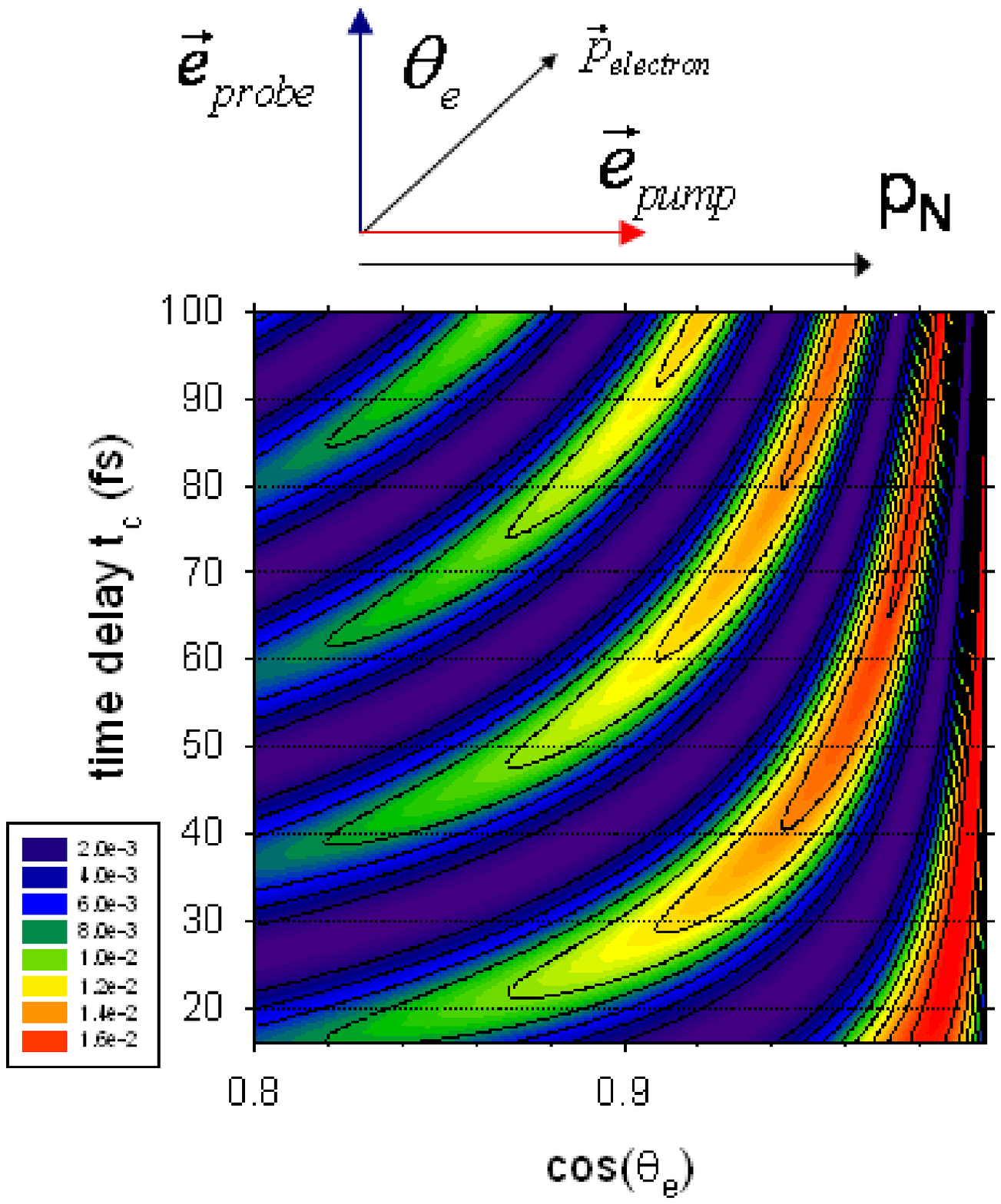}
\caption{Same as in Fig.3. but for the case when the polarizations
of the pump and probe pulses are perpendicular and the nuclear
relative momentum $\vec{p}_N$ is also parallel to the polarization
of the as shown in the upper part of the figure.}
\end{figure}
Assuming that initially \hpl was at rest with the initial momentum
of \hpl centre of mass $\vec{P}_{CM}$ is zero, we have the
following relations between the measured final momenta of the two
protons $\vec p_1$, $\vec p_2$ and of the electron momentum $\vec
p_e$ and the relative nuclear momentum $\vec p_N$:
\begin{equation}
\vec p_1 + \vec p_2 + \vec p_e =\vec{P}_{CM}=0 \;\;\;\vec p_N
=\frac{1}{2}(\vec p_1 - \vec p_2)
\label{momenta}
\end{equation}
Consequently, if the initial molecular temperature is zero it will
be sufficient to measure the electron momentum and the momentum of
one proton $\vec{p}_1$ in order to determine the $\vec{p}_N$
vector on which the interference relies. Thus the vectors present
in (\ref{probab}) formulas become:
\begin{equation}
\vec p_N=\vec p_1+\vec p_e /2 \;\;,\;\; \vec{p}_{+}=\vec p_1+\vec
p_e\;,\;\vec{p}_{-}=\vec p_1 \;. \label{pone}
\end{equation}
In the case of nonzero temperature of initial \hpl translational
motion one should either average our formula over thermal momenta
of \hpl or measure in coincidence the momenta of all three
fragments resulting from the photoionization of dissociating \hpl
in order to avoid possible washing out of the interference term.

Summarizing, our most important result is that the Fano two-centre
interference shows up in the cross term $C({p}_{+},{p}_{-},t_c)$
in eq.(\ref{interfc}) via
\begin{equation}
 \cos \left(\vec p_e \cdot  \vec{R}_N(t_c,\vec{p}_N)
\right)\;
\label{summar}
\end{equation}
where $\vec{R}_N(t_c,\vec{p}_N)$ is given in eq.(\ref{rtcappr}).
Note, that the
calculations in which $\vec{R}$ is fixed
lead instead to the  very similar interference term $\cos(\vec p_e
\cdot \vec R)$ in eq.(\ref{fixed}). Thus the effect of nuclear motion
consists in replacing $\vec R$ by $\vec{R}_N (t_c,\vec{p}_N)$
which is a simple linear function of the final relative momentum of
outgoing protons $\vec{p}_N$ and of the time delay $t_c$.
A convenient way to analyze this interference in the case of $\vec
p_N$ fixed and parallel to the probe polarization $\vec e_{probe}$
(note then $\theta_e=\theta_{pe}$ which simplifies significantly
eq.(\ref{fanopn}) is to expand the angular distributions described
by (\ref{fanopn}) in Legendre polynomials $P_l(\cos \theta_e)$:
\begin{equation} |A_{fi}|^{2} \sim \cos^{2}\theta_{e}[1-\cos(p_{e}R
\cos\theta_{e})]=\cos^{2}
\theta_{e}\left[1-\sum_{l=0,l-even}^{\infty}(2l+1)j_{l}
\left(p_{e}R(t_{c})\right)P_{l}(\cos\theta_{e})i^{l}\right]
\end{equation}
\begin{equation}
|A_{fi}|^{2}\sim
\beta_{0}(t_{c})+\beta_{2}(t_{c})P_{2}(\cos\theta_{e})+\beta_{4}(t_{c})P_{4}
(\cos\theta_{e}),\mbox{
where}
\label{betaexp}
\end{equation}
\begin{equation}
\beta_{0}=\frac{1}{3}\left[1-j_{0}+2j_{2}\right]\simeq
\frac{1}{3}\left[1-3\frac{\sin(p_{e}R_N(t_c,p_N))}{p_{e}R_N(t_c,p_N)}\right]\mbox
{   for }p_{e}R(t_{c})\gg1
\end{equation}
\begin{equation}
\beta_{2}=\frac{1}{3}\left[2-j_{0}+\frac{55}{7}j_{2}-\frac{36}{7}j_{4}
\right]\simeq\frac{1}{3}\left[2-\frac{92}{7}
\frac{\sin(p_{e}R_N(t_c,p_N))}{p_{e}R_N(t_c,p_N)}\right]\mbox
{   for }p_{e}R(t_{c})\gg1
\label{betatwo}
\end{equation}
\begin{equation}
\beta_{4}=\frac{30}{11}j_{6}-\frac{351}{77}j_{4}\simeq\frac{51}{7}
\frac{\sin(p_{e}R_N(t_c,p_N))}{p_{e}R_N(t_c,p_N)}
\mbox
{   for }p_{e}R(t_{c})\gg1 \;\;.
\end{equation}
We see clearly that the expansion of angular distributions in
Legendre polynomials reveals the Fano interference as function of
the pump-probe time delay $t_c$ in a very neat way. An analysis of
experimental data related to the Fano interference using such an
expansion was recently performed e.g. in \cite{Mabbs2005}. More
specifically, in \cite{Mabbs2005} a pump-probe experiment was
reported in which a pump pulse photodissociates a $I_2^{-}$
molecule and the probe photoionizes the dissociating molecule. The
$\beta_2 (t_{delay})$ coefficient calculated from the experimental
photoelectron angular distribution shows the modulation similar to
oscillations expected from our eq.(\ref{betatwo}). These
experimental oscillations in $\beta_2(t_{delay})$ do not survive
for the time delays larger than few picoseconds. We suggest that
this maybe related to the constant term present in
eq.(\ref{betatwo}) which shows in the experiment as background or
they disappear due to averaging over nuclear momenta which becomes
more significant at larger internuclear separations. Note, that
the higher $\beta_4(t_c)$ coefficient does not contain any
constant term and thus may yield a better contrast allowing the
Fano interference to survive for larger time delays.
\section{Some specific examples of the proposed pump-probe
experiments} The interference expected from the theory presented
in the previous section will show up most clearly when the proton
and the electron momenta are measured in coincidence. Using our
exact expressions (\ref{probab}) for probabilities as function of
the momenta of three outgoing fragments we calculate the
probabilities of ionization by the probe pulse for three selected
geometries and plot the results in Figs.3-6. Note that we do not
use in Figs.3-5 the approximations suggested in formulas
(\ref{pmapprox}),(\ref{approxa}). The probabilities are shown as
functions of the time delay $t_c$ between the pump and the probe
pulse.
We plot in Fig.3 the angular distributions of the electron in the
parallel case, i.e. all three vectors $\vec e_{pump}$, $\vec
{e}_{probe}$, and $\vec p_{N}$ are parallel and the momenta $p_e$
and $p_N$ are fixed at their maximal values. In Fig.4 the
perpendicular geometry is used, i.e. we choose the case of the
pump laser polarization $\vec e_{pump}$ perpendicular to
polarization of the probe pulse $\vec e_{probe}$. In Fig. 6 again
the geometry is parallel as in Fig.5 but instead of angular
distributions we plot there the electron spectra for the electron
flying along the polarization vectors. All three graphs show
strong interference structures as function of the time delay,  as
expected from the approximate factor
$\sin^{2}\left(\vec{p}_{e}\cdot\frac{\vec{p}_{N}}{|\vec{p}_{N}|}R_N(t_{c},p_N)
/2\right)\;.$ Note that we show in Fig.3 and in Fig.5 the
positions of the centre of wave packet corresponding to certain
time delays calculated using eq.(\ref{rtcapprab}). Similar
interference structures appear in the proton spectra displayed in
Fig.6 in which we are showing spectra as function of the single
proton momentum $p_1=|\vec{p}_1|$ with fixed electron momentum
$p_e$=0.72 a.u.. The $\vec{p}_N$ vector is calculated using
eq.(\ref{pone}).
\begin{figure}[!hb]
\includegraphics[width=0.8\textwidth]{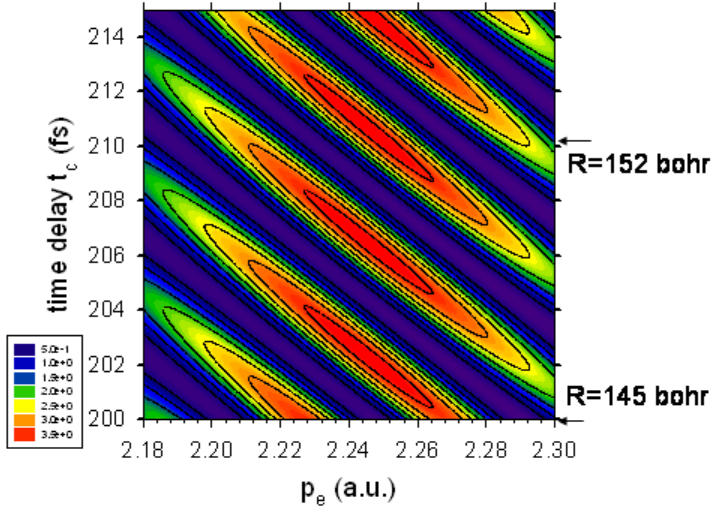}
\caption{Same as in Fig.3. but for the shorter pump wavelength
$\lambda_{probe}$=15 nm, shorter pulse duration $\tau_{FWHM}$=0.24
fs and smaller width of the wave packet $\Delta R$=1.0 bohr. Note
that now the electron angle $\theta_e$ is fixed and equal to zero.
The ionization probability is plotted now as function of the
electron momentum $p_e$.}
\end{figure}
\begin{figure}[!hb]
\includegraphics[width=0.8\textwidth]{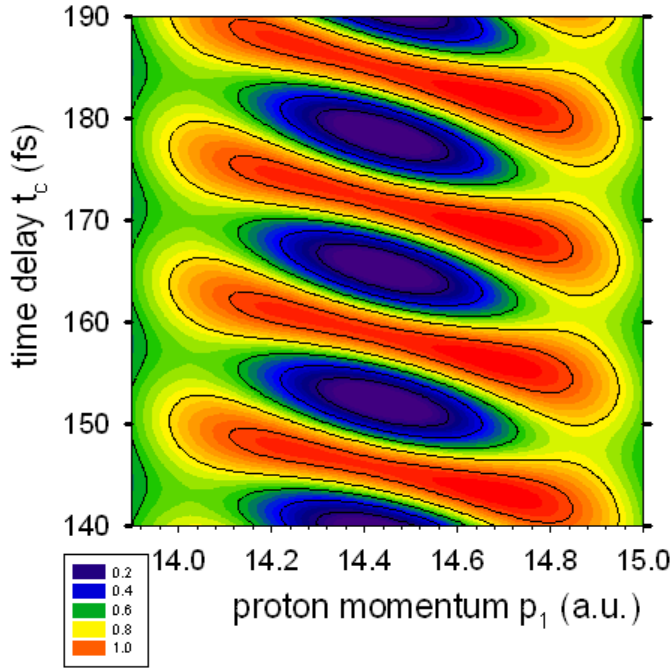}
\caption{Same as in Fig.3. but now the ionization probability is
plotted now as function of the proton momentum $p_1$ at fixed
electron angle $\theta_e=0$}
\end{figure}
\section{Concluding remarks}
Summarizing, we have investigated a laser pump-probe scheme in
which the measurement of the momenta distributions of the
photelectron  allow to determine the nuclear trajectory R(t) which
simply plays a role of a slit separation in this double-slit like
experiment in which the molecule is "illuminated from within"
\cite{Vrakking2009}. Thus the probe pulse prepares the electron source
whose de Broglie wave create interference structure. The observed
modulation as function of time delay, due to the change of the
slit-separation becomes a witness of the simultaneous presence of
the electron on each proton. In other words, when the internuclear
distance R is much larger that the bohr orbit it would be natural
to expect a localized electron on one heavy proton but such a
localization would prevent the interference seen in the
photoionization signal.

Another unusual feature of the proposed experimental scheme
in this paper is the fact
that it allows to measure, in a sense, the sign of the spatial
wave function, more specificallly the measurment we propose
detects (in the case of an ungerade initial state) the fact
that the sign of electronic wave function on one centre is minus
an an another remote centre is plus. Namely, at large
internuclear distance the electronic wave
function is given by :
$$\varphi_{el}^{g/u}=\frac{1}{\sqrt{2}}\left[\psi_{H}(\vec{r}_{e}+\vec{R}/2)
\pm \psi_{H}(\vec{r}_e-\vec{R}/2)\right]\;\;.$$ Thus the probability density
distribution $|\varphi_{el}^{g/u}(z)|^2$ is identical for both parities at very large
R-values, see Fig.2. Nevertheless, measuring the photoelectron
signal allows to distinguish between the gerade and ungerade case
since the photoionization probability at fixed internuclear
distance R is proportional to $\cos^2(\vec p_e\cdot \vec R/2)$ in
the case of dissociation occurring on the gerade electronic state
whereas is proportional to $\sin^2(\vec p_e\cdot \vec R/2)$ in the
ungerade case. This sensitivity of the photoelectron spectra
to parity (gerade or ungerade) is very specific to this simple dissociation
process and because of this feature this experimental scheme is very
distinct form the electron two-slit diffraction.

There exists already some experimental evidence for the existence
of the Fano interferences originating from dissociating molecules
at large internuclear separations: in the pump-probe experiment by
Sanov et al \cite{Mabbs2005} negative iodine $I_2^-$ ions were
used in which similar to dissociating \hpl electron delocalization
occur when a following pump-probe is used. The photoionization was
initialized using a 780-nm laser that dissociated the $I_2^-$ ions
into $I^-$ + neutral iodine atom I. After a variable time delay, a
photoionizing probe removed the electron from the $I_2^-$ ion.
Next the $\beta_2(t_{delay})$ coefficient, defined in our
eqs.(\ref{betaexp}),(\ref{betatwo}) was calculated from the
phototelectron angular distributions. This coefficient oscillates
as function of the time delay between the probe and pump pulses as
expected from eq.(\ref{betatwo}). This oscillation is due to the
fact that as in our scheme one cannot distinguish whether the
phototelectron originates from the right or left iodine atom
separated by an internuclear distance as large as 60 bohr.

Another method for the observation of the two-centre interference
was proposed in \cite{Bykov2003}. This method uses the probe pulse
which does not ionizes the dissociating molecule but is based on
elastic (Thomson) photon scattering  from the two centres in the
dissociating \hpl . Thus this method, which resembles the
double-slit experiment for photons, allows to probe the entangled
state in dissociating \hpl on the internuclear distance larger
than the method discussed in our paper since the wavelength 800 nm
laser is much larger than de Broglie wavelength used in our
scheme. This method, as ours, shows the delocalization of the
electron on two remote centres since both the Thomson scattering
or phoionization (in the case of our method) rely
on the presence of the electron on each centre, i.e neither Thomson scattering
nor photonization can
occur simply on bare proton. This fact distinguishes  both schemes
(ours and that proposed in \cite{Bykov2003}) from a simple
double-slit quantum effect.\\ \\
\appendix
\section{Beyond the Born-Oppenheimer approximation in the initial
state}

The Jacobi coordinates $\vec{r}_e,\;\vec R$ used in the section II
are not convenient at large internuclear distances R since the
wave functions $\exp (i\vec{p}\cdot \vec
R)\varphi_{H}^{u}(\vec{r}_e \pm \vec R)$ used in
eq.(\ref{initialbo}) are not the exact eigenstates of the
molecular Hamiltonian even if one neglects the Coulomb repulsion
and the attraction from the remote bare proton. These states are
approximate eigenstates, within the Born-Oppenheimer
approximation. Note, that when a proton and an hydrogen atom are
very far apart the exact eigenstates of $H_0$ are simply a product
of two plane waves for: a free proton motion, free motion of the
centre of mass of a hydrogen atom multiplied by the electronic
wave function describing the 1s electronic state of the hydrogen
atom.
Thus when the proton and the hydrogen atom are far apart and they
move with relative momentum $\vec p$ it is convenient to rewrite
the Hamiltonian $H_0$ using
different Jacobi coordinates from that used in
section II. Namely, instead of using the internuclear vector $\vec
R$ and the electron coordinate $\vec {r}_{e}$ we now use
the vectors
\begin{equation}
\vec R_1=\vec R + \alpha \vec r_{1e} \;\;\mbox{where} \;\; \vec
r_{1e}=\vec{r}_{e} - \vec{R}/2 \;\; \mbox{and}\;\;\alpha={m_e
\over m_e+m_p}.
\label{alfa}
\end{equation}
$\vec R_1$ is the relative vector between the centre of mass of a
hydrogen atom and the neighboring proton, $\vec r_{1e}$ is the
relative vector between the proton and the electron. In these
coordinates the Hamiltonian of \hpl  has the following form : (in
atomic units, $\hbar=m_e=e$=1)
\begin{equation}
\hat{H}_{0}= -{1 \over 2m''_e} \Delta_{\vec r_{1e}}-{1 \over
2\mu''}\Delta_{\vec R_1} -{1 \over |\vec r_{1e}|}
+V_2(\vec r_e,\vec R) \;\;\mbox{where}\; V_2=
-{1 \over |\vec{r}_e+\vec{R}/2|}+{1 \over R}, \label{hamilt2}
\end{equation}
where
\begin{equation}
\mu''={m_p (m_p+m_e) \over 2m_p+ m_e}\;,\;m''_e =\; {m_p m_e \over
m_p+ m_e}\;,
\end{equation}
and $m"e$ is simply the reduced mass of the electron in the
hydrogen atom. Note that as in Jacobi coordinates used in section II there is no
cross gradient term with coupling nuclear and electronic variables.
We easily find the exact eigenstates of a dissociating wave packet with
the relative momentum $\vec p $, in these new Jacobi coordinates,
in the limit of very large internuclear distance when the
potential $V_2(\vec{r}_{e},\vec{R})$ can be neglected. This exact
asymptotic eigenstate has the following form:
\begin{equation}
\psi_{in}^{asym}(\vec r_e,\vec R, \vec
p)=\exp(i\vec{p}\cdot\vec{R}_1)
\psi_H(\vec{r}_{1e}))=\exp(i\vec{p}\cdot \left(\vec{R}+\alpha
\vec{r}_{1e})\right) \psi_H(\vec r_e - \vec R/2)\;\;,
\label{initialasym}
\end{equation}
where $\vec p $ is the momentum of the relative motion between the
hydrogen atom and the remote proton. Note that in contrast to the
previously used eigenstate (\ref{initialbo}) the above state does
not have the inversion symmetry with respect to the inversion
$\vec r_e \rightarrow -\vec r_e$. Since we expect that the initial
state should have such an inversion symmetry (as being prepared
via one photon excitation of the gerade electronic state of a \hpl
molecule), we construct the ungerade initial state in the
following way
\begin{equation}
\psi_{in}={1\over \sqrt{2}}\int d^{3}p\varphi_{N}(\vec
p,R_0)\left[\psi_{in}^{asym} (-\vec r_e,\vec R, \vec
p)-\psi_{in}^{asym}(\vec r_e,\vec R, \vec p) \right]\;\;.
\label{psinapp}
\end{equation}
Clearly, for each fixed $\vec{p}$ this is an entangled state of
the two particles: a free proton and a free hydrogen atom. We
rewrite the interaction potential (\ref{vint}) in a slightly
different form :
\begin{equation}
\hat{V}_{int}= -i\vec A(t)\cdot \left({1\over m_e}\nabla_{\vec{r}_{e}}
-i{1 \over m_p} \nabla_{\vec{r}_{1p}}
-i{1 \over m_p} \nabla_{\vec{r}_{2p}}\right)=
-i{1 \over m''_ec} \vec A(t) \cdot \nabla_{\vec{r}_{1e}}
-i{1\over m_p} \vec A(t)\cdot \nabla_{\vec{r}_{2p}}
\;\;
\label{vintapp}
\end{equation}
Note that we keep here the complete interaction of the laser field
$\vec A(t)$ with three charges whereas in the previous Jacobi
coordinates (\ref{vint}) the interaction term with coupling to the
system centre of mass was not included. In the last term in the
above formula we have merged together the interaction of the
electron with the proton which binds the electron. The last term
in eq.(\ref{vintapp} will be neglected in the matrix element
(\ref{amplitude}) since it describes the interaction of the bare
proton in the case when we evaluating the term containing
$\psi_H(\vec r_{1e})$, and vice versa for the term with
$\psi_H(\vec r_{2e})$. Inserting this new initial state
(\ref{psinapp}) into the eq.(\ref{amplitude}) we get
\begin{equation}
\tilde{A}_{fi}=\tilde{N}_2 \left[A_{H}(\vec{p}_{e} +\alpha \vec {p}_{-} )\tilde{a}(\vec{p}_{-})
-A_{H}(\vec{p}_{e}-\alpha \vec {p}_{+} )\tilde{a}(\vec{p}_{+})\right]
\label{amplgenerapp}
\end{equation}
where
\begin{equation}
\tilde{a}(\vec p)=\exp [i\tilde{f}(\vec p) t_{c}-\tau^{2}\tilde{f}(\vec{p})/2]
\varphi_{N}(\vec{p},R_0) \;\;,
\label{adefapp}
\end{equation}
and
\begin{equation}
\tilde{f}(\vec p)={\vec {p}\;^2_e \over 2m'_e}+{\vec {p}\;^2_N \over
2\mu}+I_p-\Omega_{probe} -{\vec {p}\;^2 \over 2\mu''}\;\;,
\tilde{N}_2= (2\pi)^2{A_0 \tau \over 2^{3/2} m''_e  c}\;,\;
\end{equation}
We note that using exact non-Born-Oppenheimer asymptotic (for
large R) states leads to the two following modifications in the
transition amplitude as compared with (\ref{amplgeneral}). First,
the atomic transition amplitude does not factorize, second, the
function $f(\vec p)$ contains different reduced mass $\mu" $. The
first modification has a simple interpretation: since the nuclei
move in the opposite direction the relative electron momentum is
different on each centre (proton). Since the atomic amplitude
$A_{H}$ is now different at each centre previous
eqs.(\ref{probab})-(\ref{interfc}) will be modified in the
following way:
\begin{equation}
|\tilde{A}_{fi}(\vec{p}_e,\vec{p}_N,\tau,\Delta R,t_c)|^2=
 |A_H(\vec{p}_e +\alpha \vec {p}_{-} )|^2 |\tilde{a}(\vec {p}_{-}|^2 +
 |A_H(\vec{p}_e -\alpha \vec {p}_{+} )|^2 |\tilde{a}(\vec {p}_{-}|^2
+\tilde{C}({p}_{+},{p}_{-},t_c)
\label{probabapp}
\end{equation}
where
\begin{equation}
\tilde{C}({p}_{+},{p}_{-},t_c)=2|a(\vec {p}_{+})|\;|a(\vec {p}_{-})|
|A_H(\vec{p}_e +\alpha \vec {p}_{-} )|
|A_H(\vec{p}_e -\alpha \vec {p}_{+} )|
\cos (\tilde{\Phi}(t_c,))\;\;,
\label{interfcapp}
\end{equation}
\begin{equation}
\tilde{\Phi} (t_c,\vec{p}_e,\vec{p}_N)= (|\vec
{p}_{+}|-|\vec{p}_{-}|)\;R_0 +\varphi_H(\vec{p}_{e}+\alpha \vec
{p}_{-}) -\varphi_H(\vec{p}_{e}-\alpha \vec {p}_{+})\;+\; \vec{p}_e
\cdot {\vec{p}_N\over \mu''}\;t_c\;\; \label{Phinn}
\end{equation}
where the phase $\varphi_H(\vec{p})$ is simply the phase of the
atomic amplitude $A_H$ defined in eq.(\ref{atomic}), i.e.
\begin{equation}
A_H(\vec{p})=|A_H(\vec{p})|\exp [i\varphi_H(\vec{p})]\;\;.
\label{atomicn}
\end{equation}
We conclude that the Fano interference will be similar when the
non-Born-Oppenheimer correction is included. The only change in
the time delay dependent part is a replacement of the reduced mass
$\mu$ by the mass $\mu''$. Another change, due to shift of the
argument in the $A_H(\vec{p})$ function modifies only the term
which does not depend on the time delay $t_c$. Moreover the
modifications discussed in this section will be negligible for the
cases studied in section III where the values of the electron and
nuclear momenta are $p_e=0.72$, $p_N=$14.8 a.u., respectively, and
$\alpha=1/1836 $ is indeed small. Thus we do not expect that the
shifts in a slowly varying function $A_H(\vec{p})$ will modify
significantly the predictions relative to our "dynamic" Fano
interference factor (\ref{interfc}) and illustrated in previous
sections in Figs.3-6.

\acknowledgments{We gratefully acknowledge stimulating discussions
with C.L. Cocke, M. Vrakking, A. Sanov and C.-D. Lin.}

\end{document}